\documentclass[12pt]{article}
\usepackage[dvips]{graphicx}
\usepackage{epsfig}
\usepackage[section]{placeins}
\def\bea{\begin{eqnarray}}
\def\eea{\end{eqnarray}}
\def\beq{\begin{equation}}
\def\eeq{\end{equation}}

\relax

\def    \hepph  #1 {{\tt hep-ph/#1}}
\def    \hepex  #1 {{\tt hep-ex/#1}}

\newsavebox\tmpfig

\begin{document}

\pagestyle{empty}

\begin{flushright}

{\tt hep-ph/0609073}\\IFUM-873/FT\\
\end{flushright}

\begin{center}
\vspace*{0.5cm}
{\Large \bf Threshold resummation of Drell-Yan rapidity distributions
} \\
\vspace*{1.5cm}
Paolo Bolzoni \\
\vspace{0.6cm}  {\it
Dipartimento di Fisica, Universit\`a di Milano and
INFN, Sezione di Milano,\\
Via Celoria 16, I-20133 Milano, Italy}\\
\vspace*{1.5cm}
{\bf Abstract}
\end{center}

\noindent

We present a derivation of the threshold resummation formula for the
Drell-Yan rapidity distribution. Our argument is valid for all values of
rapidity  and to all orders in perturbative QCD and can be applied to all Drell-Yan processes in a universal way, i.e. both for
the production of a virtual photon $\gamma^{*}$ and the production of a vector boson $W^{\pm}$, $Z^{0}$. We
show that for the fixed-target
experiment E866/NuSea used in current parton fits, the NLL resummation corrections are comparable to NLO fixed-order
corrections and are crucial to obtain agreement with the data.
\vspace*{1cm}

\vfill
\noindent

\begin{flushleft} September 2006 \end{flushleft}
\eject

\setcounter{page}{1} \pagestyle{plain}

In perturbative QCD, it is well known that, when one approaches to the boundary of the phase space,
the cross section receives logarithmically-enhanced contributions at all
orders. These large terms have been resummed a long time ago for the classes of inclusive hadronic processes
of the type of deep-inelastic and Drell-Yan \cite{s,ct} to
next-to-leading-logarithmic order (NLL). More recently, the next-to-next-to-leading-logarithmic (NNLL)
accuracy has been reached \cite{v}. Threshold resummation of inclusive
processes can affect significantly cross sections and the extraction of parton densities \cite{mc1,mc2}.
For the case of small transverse momentum distributions in Drell-Yan processes,
it has been shown that resummation is necessary to reproduce the correct behavior of the cross section \cite{css}.

The differential rapidity Drell-Yan cross section is used for the extraction of the ratio
$\bar{d}/\bar{u}$  of parton densities. The accurate knowledge of these functions is
needed to study Higgs boson production and the asymmetry $W^{\pm}$.
The resummation of Drell-Yan rapidity distributions
was first considered in 1992 \cite{ls}. At that time, it was suggested a
resummation formula for the case of zero rapidity. Very recently, thanks to the analysis of the full NLO calculation of the Drell-Yan
rapidity distribution, it has been shown \cite{mv}, that the result given in \cite{ls} is valid at NLL
for all rapidities.

In this Letter, we will give a simple proof of an all-order
resummation formula valid  for all values of rapidity. To do this,
we will use the technique of the double Fourier-Mellin moments
developed in \cite{sv}. In particular, we will show that the
resummation can be reduced to that of the rapidity-integrated
process, which is given in terms of a dimensionless universal
function for both DY and $W^{\pm}$ and $Z^{0}$ production, and has
been largely studied \cite{s,ct} even to all logarithmic orders
\cite{fr}. Finally, we implement numerically the resummation
formula and give predictions of the full rapidity-dependent NLL
Drell-Yan cross section for the case of the fixed-target
E866/NuSea experiment. We find that resummation at the NLL level
is necessary and that its agreement with the experimental data is
better than the NNLO calculation \cite{admp}.

We consider the general Drell-Yan process in which the collisions of two hadrons ($H_{1}$ and $H_{2}$)
produce a virtual photon $\gamma^{*}$ (or an on-shell vector boson $V$) and any collection of
hadrons (X):
\bea
H_{1}(P_{1})+H_{2}(P_{2})\rightarrow \gamma^{*}(V)(Q) +X(K).
\eea
In particular, we are interested in the differential cross section $\frac{d\sigma}{dQ^{2}dY}(x,Q^{2},Y)$,
where $Q^{2}$ is the invariant mass of the photon or of the vector boson, $x$ is defined as usual as the
fraction of invariant mass that the hadrons transfer to the photon (or to the vector boson) and $Y$ is the rapidity of
$\gamma^{*}$ ($V$) in the hadronic center-of-mass:
\bea\label{rap}
x\equiv\frac{Q^{2}}{S},\quad S=(P_{1}+P_{2})^{2}, \quad Y\equiv\frac{1}{2}\ln\left(\frac{E+p_{z}}{E-p_{z}}\right),
\eea
where $E$ and $p_{z}$ are the energy and the momentum along the collisional axis of $\gamma^{*}(V)$ respectively.
At the partonic level, a parton 1(2) in the hadron $H_{1}$ ($H_{2}$) carries a longitudinal momentum $p_{1}=x_{1}P_{1}$ ($p_{2}=x_{2}P_{2}$).
Thus, the rapidity in the partonic center-of-mass ($y$) is obtained performing a boost of $Y$ between the two frames:
\bea\label{yY}
y=Y-\frac{1}{2}\ln\left(\frac{x_{1}}{x_{2}}\right).
\eea

In order to understand the kinematic configurations in terms of rapidity, it is convenient to define a new
variable $u$,
\bea\label{u}
u\equiv\frac{Q\cdot p_{1}}{Q\cdot p_{2}}=e^{-2y}=\frac{x_{1}}{x_{2}}e^{-2Y}.
\eea
which can  assume all the values in the closed interval,
\bea\label{boundary}
z\leq u\leq \frac{1}{z},
\eea
with
\bea\label{z}
z=\frac{Q^{2}}{2p_{1}\cdot p_{2}}=\frac{Q^{2}}{(p_{1}+p_{2})^{2}}=\frac{x}{x_{1}x_{2}}.
\eea
The upper and lower bounds in eq.(\ref{boundary}) are reached when the extra radiation is emitted collinear to the incoming
parton 1 and 2 respectively.
Eqs.(\ref{yY},\ref{u}) allow us to rewrite the relation in eq.(\ref{boundary}) as a relation for the upper and lower bounds of the partonic
center-of-mass rapidity:
\bea\label{boundy}
\frac{1}{2}\ln z\leq y\leq \frac{1}{2}\ln\frac{1}{z}.
\eea
Substituting eqs.(\ref{u}, \ref{z}) into the two
conditions $u\geq z$ and $u\leq 1/z$, we obtain the lower
kinematic bound for $x_{1}$ and $x_{2}$:
\bea
x_{1}\geq \sqrt{x}e^{Y}\equiv x_{1}^{0}, \quad x_{2}\geq \sqrt{x}e^{-Y}\equiv x_{2}^{0}
\eea
and the obvious requirement $x_{1(2)}^{0}\leq 1$ implies that the hadronic rapidity has a lower
and an upper bound:
\bea\label{boundY}
\frac{1}{2}\ln x\leq Y\leq \frac{1}{2}\ln\frac{1}{x}.
\eea
The variable $z$ in eq.(\ref{z}) can be viewed as the fraction of invariant mass that the incoming partons transfer to $\gamma^{*}(V)$.
and, hence, the threshold limit is reached when $z$ approaches to $1$.

According to the standard factorization of collinear singularities of perturbative QCD, the expression for the hadronic differential
cross section in rapidity has the form,
\bea\label{hadrcs}
\frac{d\sigma}{dQ^{2}dY}=\sum_{i,j}\int_{x_{1}^{0}}^{1}dx_{1}\int_{x_{2}^{0}}^{1}
dx_{2}F^{H_{1}}_{i}(x_{1},\mu^{2})F^{H_{2}}_{j}(x_{2},\mu^{2})
\frac{d\hat{\sigma}_{ij}}{dQ^{2}dy}\left(x_{1},x_{2},\frac{Q^{2}}{\mu^{2}},\alpha_{s}(\mu^{2}),y\right),
\eea
where $y$ depends on $Y$, $x_{1}$ and $x_{2}$ according to eq.(\ref{yY}).
The sum runs over all possible partonic subprocesses, $F^{(1)}_{i}$ ,$F^{(2)}_{j}$ are respectively the parton densities of the hadron
$H_{1}$ and $H_{2}$, $\mu$ is the factorization scale (chosen equal to renormalization scale for simplicity) and
$d\hat{\sigma}_{ij}/(dQ^{2}dy)$ is the partonic cross section.
In the threshold limit the gluon-quark channels are
suppressed by powers of $(1-z)$ \cite{fr} and, so, in order to study resummation, we will consider only the quark-anti-quark contributions of the sum in eq.(\ref{hadrcs}).
These last terms are related to the same dimensionless coefficient function
$C(z,Q^{2}/\mu^{2},\alpha_{s}(\mu^{2}),y)$ through the relations
\bea\label{photondy}
x_{1}x_{2}\frac{d\hat{\sigma}^{\gamma^{*}}_{q\bar{q}'}}{dQ^{2}dy}\left(x_{1},x_{2},\frac{Q^{2}}{\mu^{2}},\alpha_{s}(\mu^{2}),y\right)=\frac{4\pi\alpha^{2}c_{q\bar{q}'}}{9Q^{2}S}
C\left(z,\frac{Q^{2}}{\mu^{2}},\alpha_{s}(\mu^{2}),y\right),
\eea
for the virtual photon vertex and
\bea\label{Vprod}
x_{1}x_{2}\frac{d\hat{\sigma}^{V}_{q\bar{q}'}}{dQ^{2}dy}\left(x_{1},x_{2},\frac{Q^{2}}{\mu^{2}},\alpha_{s}(\mu^{2}),y\right)=
\frac{\pi G_{F}Q^{2}\sqrt{2}c_{q\bar{q}'}}{3S}\delta(Q^{2}-M^{2}_{V})
C\left(z,\frac{Q^{2}}{\mu^{2}},\alpha_{s}(\mu^{2}),y\right),
\eea
for the real vector boson vertex.
Here $G_{F}$ is the Fermi constant, $M_{V}$ is the mass of the produced vector boson. The coefficients $c_{q\bar{q}'}$ are given by:
\begin{eqnarray}
c_{q\bar{q}'}&=&Q_{q}^{2}\delta_{q\bar{q}}\quad\textrm{for $\gamma^{*}$},\label{gamma}\\
c_{q\bar{q}'}&=&|V_{qq'}|^{2}\quad \textrm{for  $W^{\pm}$},\label{W}\\
c_{q\bar{q}'}&=&4[(g_{v}^{q})^{2}+(g_{a}^{q})^{2}]\delta_{q\bar{q}} \quad\textrm{for $Z^{0}$},\label{Z}
\end{eqnarray}
where $Q_{q}^{2}$ is the square charge of the quark $q$, $V_{qq'}$ are the CKM mixing factors for the quark flavors $q,q'$ and
\begin{eqnarray}
g_{v}^{q}&=&\frac{1}{2}-\frac{4}{3}\sin^{2}\theta_{W},\quad g_{a}^{q}=\frac{1}{2}\quad\textrm{for an up-type quark},\\
g_{v}^{q}&=&-\frac{1}{2}+\frac{2}{3}\sin^{2}\theta_{W},\quad g_{a}^{q}=-\frac{1}{2}\quad\textrm{for a down-type quark},\label{gv}
\end{eqnarray}
with $\theta_{W}$  the Weinberg weak mixing angle.
Thus, we are left with a dimensionless cross section of the form:
\bea\label{universal}
\sigma(x,Q^{2},Y)\equiv \int_{x_{1}^{0}}^{1}\frac{dx_{1}}{x_{1}}\int_{x_{2}^{0}}^{1}\frac{dx_{2}}{x_{2}}F_{1}^{H_{1}}(x_{1},\mu^{2})
F_{2}^{H_{2}}(x_{2},\mu^{2})C\left(z,\frac{Q^{2}}{\mu^{2}},\alpha_{s}(\mu^{2}),y\right),
\eea
where $F_{1}$ and $F_{2}$ are quark or anti-quark parton densities in the hadron $H_{1}$ and $H_{2}$ respectively. This shows the universality of resummation in
Drell-Yan processes in the sense that only the quantity defined in eq.(\ref{universal}) has to be resummed.

We shall now show that the resummed expression of eq.(\ref{universal}) is obtained by simply replacing
the coefficient function $C(z,Q^{2}/\mu^{2},\alpha_{s}(\mu^{2}),y)$ with its integral over y, resummed to the desired
logarithmic accuracy. To show this, we recall that resummation is usually performed in the space of the variable $N$, which is the Mellin
conjugate of $x$, since Mellin transformation turns convolution products into ordinary products. In the case of the rapidity
distribution, however, this is not sufficient.
In fact, we see that the Mellin transform with respect to $x$,
\bea\label{mellintrans}
\sigma(N,Q^{2},Y)\equiv \int_{0}^{1}dxx^{N-1}\sigma(x,Q^{2},Y),
\eea
does not diagonalize the double integral in eq.(\ref{universal}), because the partonic center-of-mass rapidity $y$ depends on $x_{1}$ and $x_{2}$ through eq.(\ref{yY}).
The ordinary product in Mellin space can be recovered performing the Mellin transform with respect to $x$ of the Fourier transform with respect to $Y$.
Using eqs.(\ref{boundY},\ref{boundy}) and the fact that the coefficient function must be symmetric in $y$, we find
\begin{eqnarray}\label{fact}
\sigma(N,Q^{2},M)&\equiv&\int_{0}^{1}dxx^{N-1}\int_{\log\sqrt{x}}^{\log1/\sqrt{x}}dYe^{iMY}\sigma(x,Q^{2},Y)\\
&=&F_{1}^{H_{1}}(N+iM/2,\mu^{2})F_{2}^{H_{2}}(N-iM/2,\mu^{2})C\left(N,\frac{Q^{2}}{\mu^{2}},\alpha_{s}(\mu^{2}),M\right),
\end{eqnarray}
where
\begin{eqnarray}
F_{i}^{H_{i}}(N\pm iM/2,\mu^{2})&=&\int_{0}^{1}dxx^{N-1\pm iM/2}F_{i}^{H_{i}}(x,\mu^{2}),\label{mmpdf}\\
C\left(N,\frac{Q^{2}}{\mu^{2}},\alpha_{s}(\mu^{2}),M\right)&=&2\int_{0}^{1}dzz^{N-1}\int_{0}^{\log1/\sqrt{z}}dy\cos(My)
C\left(z,\frac{Q^{2}}{\mu^{2}},\alpha_{s}(\mu^{2}),y\right)\label{mfcoefffunc2}.\label{coefftrans}
\end{eqnarray}

The dependence on $M$, the Fourier conjugate of the rapidity $y$, originates from the parton
densities, that depend on $N\pm iM/2$, and from the factor of $\cos(My)$ in the integrand of eq.(\ref{mfcoefffunc2}).
This last dependence, however, is irrelevant in the large-$N$ limit. Indeed, one can expand $\cos(My)$ in powers of $y$,
\bea
\cos(My)=1-\frac{M^{2}y^{2}}{2}+\emph{O}(M^{4}y^{4}).\label{expansion}
\eea
and observe that the first term of this expansion leads to a convergent integral (the rapidity-integrated cross section), while
the following terms are suppressed by powers of $(1-z)$, since the upper integration bound is
\bea
\ln\frac{1}{\sqrt{z}}=\frac{1}{2}(1-z)+\emph{O}((1-z)^2).
\eea
Hence, up to terms suppressed
by factors $1/N$, eq.(\ref{mfcoefffunc2}) is equal to the Mellin transform of the rapidity-integrated Drell-Yan coefficient function that we call
$C_{I}(N,Q^{2}/\mu^{2},\alpha_{s}(\mu^{2}))$.
This completes our proof. We get
\bea\label{rescrosssec}
\sigma^{res}(N,Q^{2},M)=F_{1}^{H_{1}}(N+iM/2,\mu^{2})F_{2}^{H_{2}}(N-iM/2,\mu^{2})C_{I}^{res}\left(N,\frac{Q^{2}}{\mu^{2}},\alpha_{s}(\mu^{2})\right).
\eea

This is the main theoretical result of our Letter: it shows that,
near threshold, the Mellin-Fourier transform of the coefficient
function does not depend on the Fourier moments and that this is
valid to all orders of QCD perturbation theory. Furthermore this
result remains valid for all values of hadronic center-of-mass
rapidity, because we have introduced a suitable integral transform
over rapidity. The resummed rapidity-integrated Drell-Yan
coefficient function to NLL is well known \cite{s,ct} and, using
the notation of \cite{fr}, it is given in a compact form (in the
$\overline{MS}$ scheme) by \bea\label{rescoeff}
C_{I}^{res}\left(N,\frac{Q^{2}}{\mu^{2}},\alpha_{s}(\mu^{2})\right)=\exp\left\{-\int_{1}^{N^{2}}\frac{dn}{n}
\left[\int_{n\mu^{2}}^{Q^{2}}\frac{dk^{2}}{k^{2}}\left(A_{1}\alpha_{s}(\frac{k^{2}}{n})+A_{2}\alpha_{s}^{2}(\frac{k^{2}}{n})\right)+B_{1}\alpha_{s}(\frac{Q^{2}}{n})\right]\right\},
\eea where \bea\label{coeff} A_{1}=\frac{C_{F}}{\pi},\quad
A_{2}=\frac{C_{F}}{2\pi^{2}}\left[C_{A}\left(\frac{67}{18}-\frac{\pi^{2}}{6}\right)-\frac{5}{9}N_{f}\right],\quad
B_{1}=-\frac{\gamma_{E}A_{1}}{2\pi} \eea with $C_{F}=4/3$,
$C_{A}=3$, $N_{f}$ the number of flavors and with the Euler gamma
$\gamma_{E}=0.5772\dots$ . The use of only the first coefficient
$A_{1}$ allows us to resum all the LL contributions
$\alpha_{s}^{k}\log^{k+1}(N)$ while the use of all the three
coefficients in eq.(\ref{coeff}) enable us to add also the NLL
terms $\alpha_{s}^{k}\log^{k}(N)$.

A NLL expression of the rapidity distribution is obtained by taking the inverse Mellin and Fourier transform of $\sigma^{res}(N,Q^{2},M)$. This procedure requires
the use of some specific prescription \cite{cmnt,lsn,frru} in order to overcome the problem of the Landau singularity in $\alpha_{s}(Q^{2}/N^{2})$.
Here, we adopt the ``Minimal Prescription'' proposed in \cite{cmnt}, which is simply obtained choosing the integration contour of the inverse Mellin transform in such a way
that all the poles
of the integrand are to the left, except the Landau pole. Furthermore,
in order to improve numerical convergence and to avoid the singularities of the parton densities of eq.(\ref{rescrosssec}) which are computed out of the real axis,
we perform the $N$-integral along a path $\Gamma$ given by:
\begin{eqnarray}
\Gamma&=&\Gamma_{1}+\Gamma_{2}+\Gamma_{3}\\
\Gamma_{1}(t)&=&C_{MP}-i\frac{M}{2}+t(1+i),\quad t\in (-\infty,0)\\
\Gamma_{2}(s)&=&C_{MP}+is\frac{M}{2},\quad s\in (-1,1)\\
\Gamma_{3}(t)&=&C_{MP}+i\frac{M}{2}-t(1-i),\quad t\in (0,+\infty)
\end{eqnarray}
where $C_{MP}$ is a positive number below the Landau pole of $\alpha_{s}(Q^{2}/N^{2})$.
Performing the changes of variable $M=-\ln m$ and $t=-\ln s$, the double inverse transform over the curve $\Gamma$ becomes:
\bea\label{trasff}
\sigma^{res}(x,Q^{2},Y)=\frac{1}{\pi}\int_{0}^{1}\frac{dm}{m}\cos(-Y\ln m)\sigma^{res}(x,Q^{2},-\log m),
\eea
where $\sigma^{res}(x,Q^{2},M)$  is given by
\begin{eqnarray}\label{trasfm}
&&\qquad\qquad\qquad\qquad\qquad\qquad\sigma^{res}(x,Q^{2},M)=\\
&&\frac{1}{\pi}\int_{0}^{1}\frac{ds}{s}\Re\bigg[x^{-C_{MP}-\ln s+i(M/2+1)}\sigma^{res}(C_{MP}+\ln s-i(M/2+1),Q^{2},M)(1-i)\nonumber\\
&&+\frac{sM}{2}x^{-C_{MP}-isM/2}\sigma^{res}(C_{MP}+isM/2, Q^{2},M)\bigg].\nonumber
\end{eqnarray}

Eqs.(\ref{trasff},\ref{trasfm}) are the expressions that we use to
evaluate numerically the resummed dimensionless cross section in the variables $x$ and $Y$. The explicit expression of eq.(\ref{rescoeff}) is easily obtained performing the integrals
and is given by
\bea\label{explform}
C_{I}^{res}\left(N,\frac{Q^{2}}{\mu^{2}},\alpha_{s}(\mu^{2})\right)=\exp\{\ln N g_{1}(\lambda)+g_{2}(\lambda)\},
\eea
where
\begin{eqnarray}
g_{1}(\lambda)&=&\frac{A_{1}}{\beta_{0}\lambda}[2\lambda+(1-2\lambda)\log(1-2\lambda)]\label{g1}\\
g_{2}(\lambda)&=&-\frac{2A_{1}\gamma_{E}}{\beta_{0}}\log(1-2\lambda)+\frac{A_{1}\beta_{1}}{\beta_{0}^{3}}[2\lambda+\log(1-2\lambda)+\frac{1}{2}\log^{2}(1-2\lambda)]\nonumber\\
&&-\frac{A_{2}}{\beta_{0}^{2}}[2\lambda+\log(1-2\lambda)]+\log\left(\frac{Q^{2}}{\mu^{2}}\right)\frac{A_{1}}{\beta_{0}}\log(1-2\lambda)\label{g2}
\end{eqnarray}
and
\bea
\lambda=\beta_{0}\alpha_{s}(\mu^{2})\ln N,\quad\beta_{0}=\frac{1}{4\pi}\left(11-\frac{2}{3}N_{f}\right),\quad\beta_{1}=\frac{1}{16\pi^{2}}\left(102-\frac{38}{3}N_{f}\right)
\label{betacoeff}.
\eea
Here, we choose the factorization scale equal to the renormalization scale for simplicity. To study the dependence on the renormalization scale one has simply to express
$\alpha_{s}(\mu^{2})$ in terms of it. Furthermore, we need the analytic continuations to the whole complex plane of the Mellin-transformed parton densities that appear
in eq.(\ref{rescrosssec}). In order to overcome this problem, we have to evolve up a partonic fit taken at a certain scale solving
the DGLAP evolution equations in Mellin space \cite{grv}. The LO and NLO expressions of the
splitting functions are reported in \cite{fkl} and their analytic continuations are given in \cite{gly} and \cite{bk}.

Finally, we want to obtain a NLO determination of the cross section improved with NLL resummation. In order to do this, we must keep the resummed dimensionless part of the cross
section eq.(\ref{trasff}), multiply it by the correct dimensional prefactors looking eqs.(\ref{photondy}-\ref{gv}), add the full NLO cross section and subtract the
double-counted logarithmic enhanced contributions. This matching has to be done in the $x$ and $Y$ spaces, because we are not able to calculate the
Mellin-Fourier moments of the full NLO cross section analytically. Thus, we have
\bea\label{matching}
\frac{d\sigma}{dQ^{2}dY}=\frac{d\sigma^{FO}}{dQ^{2}dY}+\frac{d\sigma^{res}}{dQ^{2}dY}-\left[\frac{d\sigma^{res}}{dQ^{2}dY}\right]_{\alpha_{s}=0}-
\alpha_{s}\left[\frac{\partial}{\partial\alpha_{s}}\left(\frac{d\sigma^{res}}{dQ^{2}dY}\right)\right]_{\alpha_{s}=0}.
\eea
The first term is the full NLO cross section reported in \cite{mv,g,smrs,aem}, which includes even the quark-gluon channel. The
third and the fourth terms in eq.(\ref{matching}) are obtained in the same way as the second one, but with the substitutions
\begin{eqnarray}
C_{I}^{res}\left(N,\frac{Q^{2}}{\mu^{2}},\alpha_{s}(\mu^{2})\right)&\rightarrow& 1,\\
C_{I}^{res}\left(N,\frac{Q^{2}}{\mu^{2}},\alpha_{s}(\mu^{2})\right)&\rightarrow& \alpha_{s}(\mu^{2})
2A_{1}\left\{\ln^{2}N+\ln N\left[2\gamma_{E}-\log\left(\frac{Q^{2}}{\mu^{2}}\right)\right]\right\}\label{matchtrm},
\end{eqnarray}
respectively.
The terms that appear in eq.(\ref{matchtrm}) are exactly the $\emph{O}(\alpha_{s})$ logarithmic enhanced contributions in the $\overline{MS}$ scheme.

We note that the final expression eq.(\ref{matching}) is relevant  even when the variable $x$ is not large.  In fact, the cross section can get the dominant
contributions from the integral in eq.(\ref{universal}) for values of $z$ eq.(\ref{z}) that are near the threshold even when $x$ is not close to one, because
of the strong
suppression of parton densities $F_{i}(x_{i},\mu^{2})$ when $x_{i}$ are large.

To show the importance of this resummation, we have calculated the Drell-Yan rapidity distribution for
proton-proton collisions at the Fermilab fixed-target experiment E866/NuSea \cite{w}.  The center-of-mass energy has been fixed at $\sqrt{S}=38.76\, \textrm{GeV}$ and the
invariant mass of the virtual photon $\gamma^{*}$  has been chosen to be $Q^{2}=64\, \textrm{GeV}^{2}$ in analogy with \cite{admp}. Clearly the
contribution of the virtual $Z^{0}$ can be neglected, because its mass is much bigger than $Q^{2}$. In this case $x=0.04260$ and the upper and lower bound of the
hadronic rapidity $Y$ eq.(\ref{boundY}) are given by $\pm 1.57795$. We have evolved up the  MRST 2001 parton distributions (taken at  $\mu^{2}=1\,\textrm{GeV}^{2}$)
as in \cite{admp}, where the NNLO calculation is performed. However, results obtained using more modern parton sets should not be very different.
The LO parton set is given in \cite{mrst2} with $\alpha_{s}^{LO}(m_{Z})=0.130$ and the NLO set is given in \cite{mrst} with $\alpha_{s}^{NLO}(m_{Z})=0.119$.
The evolution of parton densities at the scale $\mu^{2}$  has been performed in the variable
flavor number scheme. The quarks has been considered massless and, at the scale of the transition of the flavor number ($N_{f}\rightarrow N_{f}+1$), the new flavor is generated
dynamically. The resummation formula eq.(\ref{rescrosssec}) together with eqs.(\ref{explform}-\ref{betacoeff}) has been used with the number of flavors $N_{f}=4$.

\begin{figure}
\begin{center}
\includegraphics[scale=0.6]{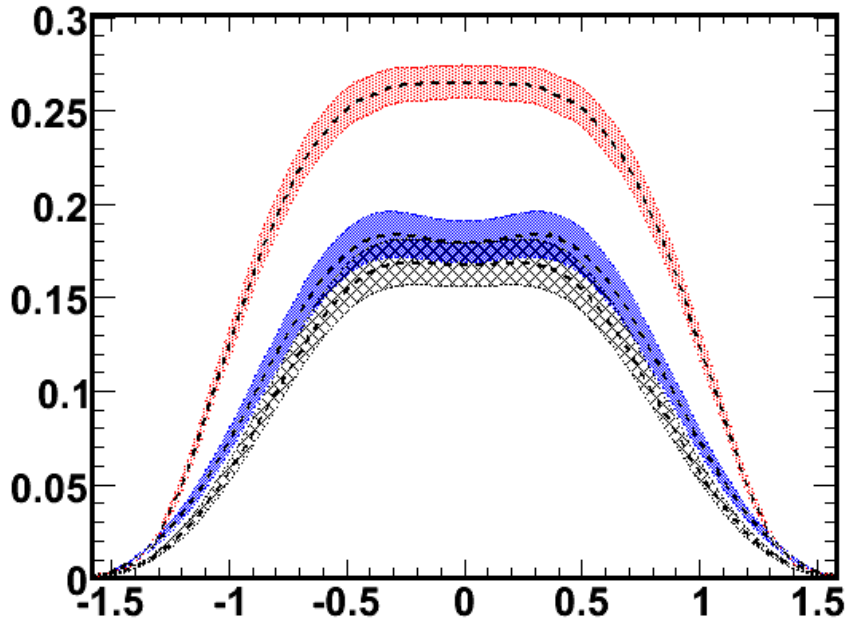}
\caption{\footnotesize{Y-dependence of $d^{2}\sigma/(dQ^{2}dY)$ in units of
$\textrm{pb}/\textrm{GeV}^2$. The curves are, from top to bottom, the NLO result (red band), the LO+LL
resummation (blue band) and the LO (black band). The bands are obtained varying
the factorization scale between $\mu^{2}= 2Q^{2}$ and $\mu^{2}=1/2 Q^{2}$.}}
\label{LOLLNLO}
\end{center}
\end{figure}

\begin{figure}
\begin{center}
\includegraphics[scale=0.6]{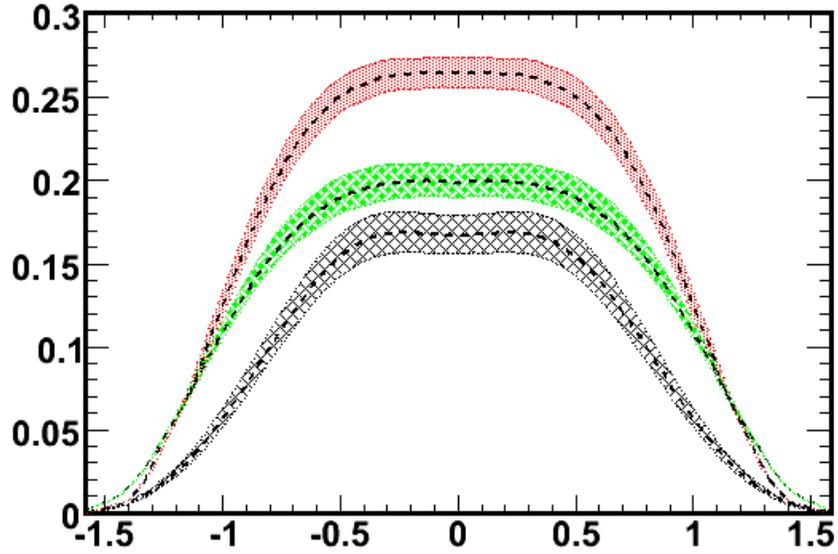}
\caption{\footnotesize{Y-dependence of $d^{2}\sigma/(dQ^{2}dY)$ in units of
$\textrm{pb}/\textrm{GeV}^2$. The curves are, from top to bottom, the NLO result (red band),
the NLO+NLL resummation (green band)
and the LO (black band). The bands are obtained as in figure \ref{LOLLNLO}.}}\label{LONLONLL}
\end{center}
\end{figure}

\begin{figure}
\begin{center}
\includegraphics[scale=0.6]{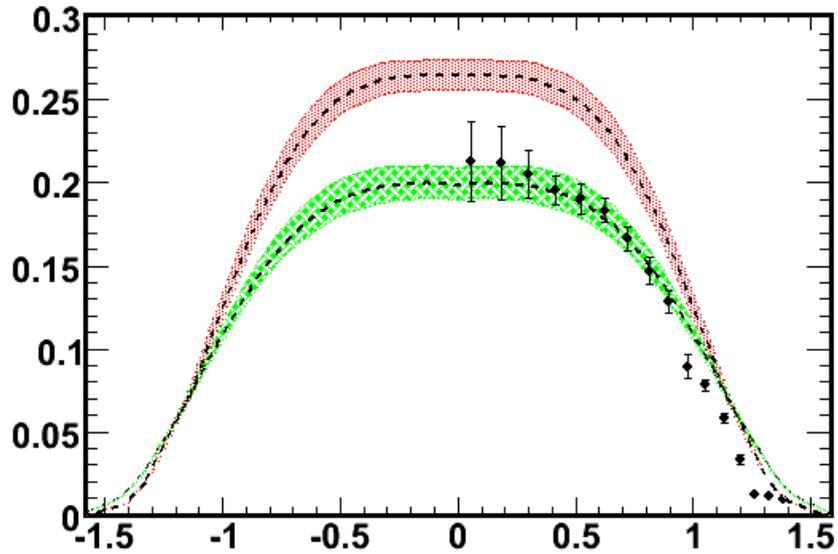}
\caption{\footnotesize{Y dependence of $d^{2}\sigma/(dQ^{2}dY)$ in units of
$\textrm{pb}/\textrm{GeV}^2$. The curves are, from top to bottom, the NLO result (red band) and the NLO+NLL resummation
(green band) together with the E866/NuSea data. The bands are obtained as in figure \ref{LOLLNLO}.}}
\label{NLONLLDT}
\end{center}
\end{figure}

In figure \ref{LOLLNLO}, we plot  the rapidity-dependence of the cross section at LO, NLO and LO improved with LL resummation. The effect of LL resummation
is small compared to the effect of the full NLO correction. We see that, at leading order, the impact of the resummation is negligible in comparison to the NLO
fixed-order correction. This means that the NLL resummation is necessary.

The LO, the NLO and its NLL improvement cross sections are shown in figure \ref{LONLONLL}.
The effect of the NLL resummation in the central rapidity region is almost as large as
the NLO correction, but it reduces the cross section instead of enhancing it for not large values of rapidity.
Going from the LO result to the NLO with NLL resummation,
we note a reduction of the dependence on the
factorization scale i.e. a reduction of the theoretical error. It is interesting to observe that logarithmically enhanced
and constant terms account for more than 80\% of the NLO contribution for all relevant rapidities. Therefore, they have the same sign. Nevertheless a
suppression arises due to the shift in the complex plane of the dominant contribution of the resummed exponent. This suppression starts at order $\emph{O}(\alpha_{s}^{2})$.

In figure \ref{NLONLLDT}, we report the experimental data of \cite{w} converted to the $Y$ variable, together with our NLO and NLL resummed predictions. The agreement
with data is good and a great improvement  for not large rapidity is obtained with respect to the NLO calculation.
We note also that the NLL resummation gives better result than the NNLO calculation performed in \cite{admp}. The NNLO prediction has a worse agreement
with data than the NLO one for not large values of rapidity.
This result suggests that, for the case of rapidity distributions, NLL resummation is more important than high-fixed-order calculation
and that it can be so even at higher center-of-mass energies.

To summarize, we have proved a resumation formula for the Drell-Yan rapidity distributions to all logarithmic accuracy and valid for all values of rapidity. Isolating a
universal dimensionless coefficient function,
which is exactly that ones of the Drell-Yan rapidity-integrated, we have shown a general procedure to obtain resummed results to NLL for the rapidity distributions of a virtual
photon $\gamma^{*}$ or of a real vector boson $W^{\pm},Z^{0}$. Furthermore, we have outlined a general method to
calculate numerical predictions and analyzed the impact of resummation for the fixed-target experiment E866/NuSea. This
shows that NLL resummation has an important effects on predictions of differential rapidity cross sections giving an agreement
with data that is better than NNLO full calculations.
\\\\
\textbf{Acknowledgements}
\\

I want to thank S.~Forte and G.~Ridolfi for suggesting this problem; A.~Vicini for useful discussions about numerical calculations;
G.~Cerati, B.M.~Mognetti, S.~Montesano and A.~Zelioli for their concern about my programming difficulties.

\end{document}